\documentclass[preprint, 
superscriptaddress,
 amsmath,amssymb,
 aps,
floatfix
]{revtex4-1}

\usepackage{graphicx}
\usepackage{dcolumn}
\usepackage{bm}
\usepackage{hyperref}
\usepackage{upgreek}
\usepackage{xcolor}
\usepackage{comment}

\begin{document}

\title{Thermodynamically Consistent Continuum Theory of Magnetic Particles in High-Gradient Fields}

\author{Marko Tesanovic\footnote[1]{These two authors contributed equally to this work.}}
\email{e-mail: m.tesanovic@tum.de}
\affiliation{Technical University of Munich, School of Engineering and Design, Chair of Bioseparation Engineering, Boltzmannstraße 15, 85748 Garching, Germany}

\author{Daniel M. Markiewitz\footnotemark[1]}
\affiliation{Department of Chemical Engineering, Massachusetts Institute of Technology, Cambridge, Massachusetts 02139 USA}

\author{Marcus L. Popp}
\affiliation{Technical University of Munich, School of Engineering and Design, Chair of Bioseparation Engineering, Boltzmannstraße 15, 85748 Garching, Germany}

\author{Martin Z. Bazant}
\affiliation{Department of Chemical Engineering, Massachusetts Institute of Technology, Cambridge, Massachusetts 02139 USA}
\affiliation{Department of Mathematics, Massachusetts Institute of Technology, Cambridge, Massachusetts 02139 USA}

\author{Sonja Berensmeier}
\affiliation{Technical University of Munich, School of Engineering and Design, Chair of Bioseparation Engineering, Boltzmannstraße 15, 85748 Garching, Germany}
\affiliation{Technical University of Munich, Munich Institute of Integrated Materials, Energy and Process Engineering (MEP), Lichtenbergstraße 4a, Garching, 85748, Germany}

\date{\today}
            
\begin{abstract}
Magnetic particles underpin a broad range of technologies, from water purification and mineral processing to bioseparations and targeted drug delivery. The dynamics of magnetic particles in high-gradient magnetic fields—encompassing both their transport and eventual capture—arise from the coupled interplay of field-driven drift, fluid advection, and particle–field feedback. These processes remain poorly captured by existing models relying on empirical closures or discrete particle tracking. Here, we present a thermodynamically consistent continuum theory for collective magnetic particle transport and capture in high-gradient fields. The framework derives from a free-energy functional that couples magnetic energy, entropic mixing, and steric interactions, yielding a concentration-dependent susceptibility via homogenization theory. The resulting equations unify magnetism, mass transport, and momentum balances without \textit{ad hoc} shut-off criteria, allowing field shielding, anisotropic deposition, and boundary-layer confinement to emerge naturally. Simulations predict canonical capture morphologies—axially aligned plumes, crescent-shaped deposits, and nonlinear shielding—across field strengths and flow regimes, consistent with trends reported in prior experimental and modeling studies. By organizing captured particle mass data into a dimensionless phase diagram based on the Mason number, we reveal three distinct regimes—thermodynamically controlled, transitional, and dynamically controlled. This perspective provides a predictive platform for \textit{in silico} optimization and extension to three-dimensional geometries, and informing digital twin development for industrial-scale high-gradient magnetic separation processes.
\end{abstract}

\maketitle

\section{Introduction}

Ranging from nanometers to microns in size, magnetic particles underpin technologies in contaminant removal for chemical processing and wastewater treatment, catalytic immobilization, bioseparations, targeted drug delivery, and molecular diagnostics~\cite{lee_numerical_2022, nithya_magnetic_2021, srivastava_magnetic_2025,schwaminger_magnetic_2019}. In all of these applications, particle transport is governed by externally applied magnetic field gradients that focus, steer, and modulate suspended particles in space and time. These dynamics determine both local accumulation patterns and macroscopic performance metrics such as efficiency, selectivity, and throughput.

A typical setting to study magnetic particle physics is high-gradient magnetic separations (HGMS), where magnetized matrix elements or permanent magnets create strong, spatially varying fields that drive particles toward localized maxima~\cite{chen_dynamic_2012}. Originally developed for mineral extraction and environmental remediation~\cite{kolm_research_1976}, HGMS now serves as a core technology in biotechnology and medicine for label-based separations under continuous flow~\cite{schwaminger_magnetic_2019, tesanovic_towards_2025}. Across these diverse implementations, the governing mechanism behind these applications is unified. First, particles are advected by the carrier fluid while simultaneously migrating along magnetic energy gradients. Then as they accumulate, they perturb both the magnetic and hydrodynamic fields, initiating nonlinear feedback loops that reshape field lines, distort streamlines, and alter capture efficiency.

This universal coupling of field, flow, and structure can be viewed with a self-consistent description via free-energy minimization.  In this perspective, particle accumulation in magnetic systems perturbs the local field and drives nonlinear restructuring of the surrounding medium, which is analogous to how induced charge reshapes electrokinetic flows~\cite{bazant_towards_2009}.

Despite widespread empirical success~\cite{ebeler_magnetic_2018, ebeler_one-step_2018, krolitzki_how_2023, brechmann_pilot-scale_2019, tesanovic_modeling_2025, turco_combined_2023}, theoretical models of HGMS remain incomplete. Classical formulations often treat flow, field, and particle phases in isolation or employ heuristic closures that fail under strongly coupled nonlinear regimes. Dynamic buildup on magnetized collectors leads to shielding, crowding, and flow obstruction. These collective effects are frequently approximated by empirical shutoff functions or saturation thresholds rather than derived from first principles.

Recent efforts have begun addressing these limitations. Hu~\textit{et al.} developed a semi-analytical framework for deposition on single wires in pulsating HGMS~\cite{hu_dynamic_2022}, modeling growth kinetics from experimental fits and assuming static fields and constant drag, thus neglecting feedback from deposit evolution. Chen~\textit{et al.} introduced a dynamic front-tracking model~\cite{chen_dynamic_2012} that couples finite-element solutions of the magnetic and flow fields with computationally intensive Lagrangian particle tracking, capturing dendritic instabilities and shielding-induced capture loss. However, this approach assumes irreversible sticking and omits steric restructuring within the deposit. Choomphon~\textit{et al.} advanced a two-phase Euler–Euler framework~\cite{choomphon-anomakhun_simulation_2017} and solved the coupled momentum balances for fluid and particle phases. To mimic crowding, they imposed empirical viscosity divergence and force shutoff functions—an effective but thermodynamically inconsistent treatment, where saturation arises from numerical limits rather than free-energy considerations. However, what remains absent is a unified continuum theory grounded in statistical thermodynamics, capable of resolving the coupled evolution of concentration, magnetic field, and flow while remaining scalable across geometries and operating conditions.

In this work, we present such a framework: a mean-field continuum model derived from a robust generalizable free-energy functional incorporating magnetic energy, entropic mixing, and an enthalpy density, in this instance using the Carnahan–Starling steric repulsion. The model introduces a concentration-dependent magnetic susceptibility consistent with homogenization bounds (e.g. Hashin–Shtrikman), enabling self-consistent feedback between particle accumulation and field distortion. The resulting equations couple magnetic potential, velocity field, and particle concentration, predicting phenomena such as shielding, self-limiting accumulation, and nonlinear front propagation without empirical shutoff criteria or discrete particle tracking. This approach unifies prior HGMS models within a rigorous statistical mechanics theory and enables predictive simulation of magnetic separations across scales—from microfluidic diagnostics to large scale separators—providing a foundation for digital-twin-based process optimization.

\section{Theory}\label{sec:theory}

Here, we consider a homogenized suspension of magnetic nanoparticles (MNP) subjected to high-gradient magnetic fields under flow. Three coupled classes of equations arising from distinct physical principles govern the system. (A) First, the thermodynamic equations capture how the magnetic field interacts with the suspension, linking magnetization, chemical potential, and stress via a free energy functional. (B) Second, mass transport equations describe the advection–diffusion of particles and the buildup of concentration gradients under flow and field. (C) Third, the hydrodynamic equations govern the momentum balance of the slurry, incorporating body forces derived from magnetic and osmotic stresses. Together, these equations form a unified continuum description where field–flow–structure coupling emerges naturally from the underlying energetics rather than through empirical closures.

\subsection{ Magnetism }
To understand suspensions of magnetic nanoparticles, one must first understand magnetism. Given that our system is a complex solution and not a simple vacuum, it is clearest to proceed with the macroscopic Maxwell's equations.
    \begin{equation}
        \nabla\cdot\textbf{D} = \rho_f
    \end{equation}
    \begin{equation}
        \nabla\cdot\textbf{B} = 0
    \end{equation}
    \begin{equation}
        \nabla\times\textbf{E} = -\frac{\partial\textbf{B}}{\partial t}
    \end{equation}
    \begin{equation}
        \nabla\times\textbf{H} = \left(\textbf{J}_f+\frac{\partial\textbf{D}}{\partial t}\right)
    \end{equation}

\noindent where \textbf{E} is the electric field, \textbf{D} is the electric displacement field, \textbf{B} is the magnetic field, \textbf{H} is the magnetic field intensity, \textbf{J}$_f$ is the free current density, $\rho_f$ is the free charge density, and $t$ is time. 

Under magnetostatic conditions with no free current density, quasi-steady fields ($\partial\textbf{D}/\partial t = 0$ and $\partial\textbf{B}/\partial t = 0$) and no free current density ($\textbf{J}_f = 0$), these electromagnetic equations connecting the electric and magnetic fields are decoupled. Furthermore, considering only the magnetic equations, they reduce to Gauss’s law for magnetism and Ampère’s law with no free current density:
    \begin{equation}
        \nabla\cdot\textbf{B} = 0
    \end{equation}
    \begin{equation}
        \nabla\times\textbf{H} = 0
    \end{equation}

\noindent Given that the curl of the \textbf{H}-field is zero, it follows that the \textbf{H}-field can be written as a scalar potential, i.e. $\textbf{H} = - \nabla\psi$. Here, $\psi$ is known as the magnetic scalar potential. Additionally, the \textbf{B}-field can be written as the sum of the magnetic \textbf{H}-field and the magnetization field (\textbf{M}-field).
    \begin{equation}
        \textbf{B} = \mu_0\left(\textbf{H}+\textbf{M}\right)
    \end{equation}

The \textbf{M}-field is a function of the \textbf{H}-field; thus, constitutive relationships are necessary to proceed. For example in ferromagnetic materials, this relationship is not direct and can display hysteresis from magnetization remaining after the field stops being applied. This effect is known as remanence. However in superparamagnetic particles, no remanence is present simplifying constitutive relationships. Additionally, magnetic materials can display magnetic saturation where increasing the externally applied \textbf{H}-field does not increase the magnetization. Generally in diamagnetic and paramagnetic materials, the common constitutive relationship is linear:
    \begin{equation}
        \textbf{M} = \chi\textbf{H}
    \end{equation}
\noindent where $\chi$ is the magnetic susceptibility that represents how strongly an applied magnetic field can magnetize a medium.
    
Therefore, under magnetostatic conditions with no free current density, a linear constitutive relationship between the \textbf{M}-field and \textbf{H}-field, and expressing the \textbf{H}-field as a scalar potential, one obtains a singular governing equation for understanding the magnetostatics from first principles:
    \begin{equation}
        -\mu_0\nabla\cdot\left(\left(1+\chi\right)\nabla\psi\right) = 0
    \end{equation}

Solving this equation works well for regions of known magnetic susceptibility and geometries. However when working with a dynamic suspension of magnetic nanoparticles that have different shapes and are aggregating, tracking everything and solving this equation becomes computationally restrictive. This restriction motivates a thermodynamically consistent continuum approach to model inhomogeneous magnetic suspension that can be achieved with two elements. First, we will propose a macroscopic constitutive equation for the magnetic susceptibility variations' dependence on the concentration of magnetic nanoparticles and the \textbf{H}-field. Second, we will derive from a free energy functional the mean-field thermodynamic equations that capture how the magnetic field interacts with the suspension, linking magnetization, chemical potential, and stress in the inhomogeneous magnetic suspension.

\subsubsection{ Macroscopic magnetic susceptibility }

\begin{figure}[ht]
    \centering
    \includegraphics[width=0.7\linewidth]{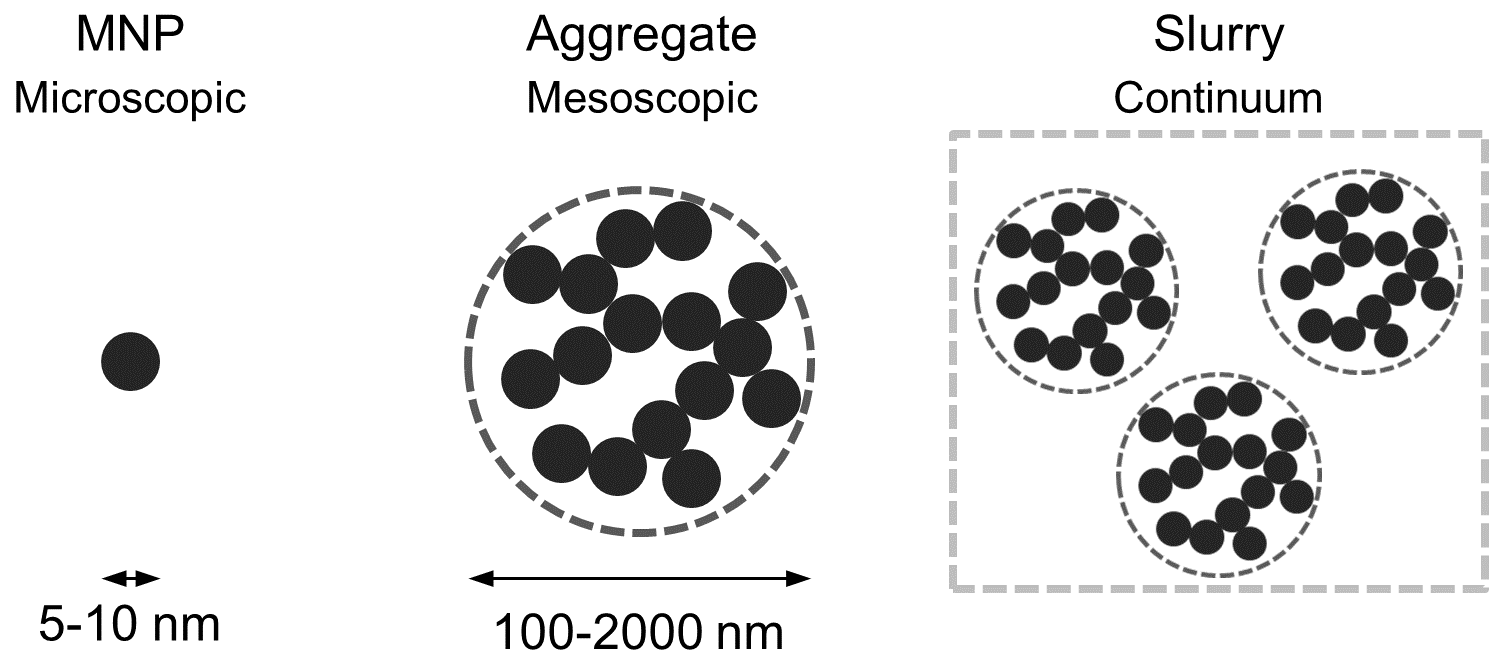}
   \caption{
    Hierarchical structure of the magnetic nanoparticle (MNP) suspension. 
    (\textbf{Microscopic}) Single MNPs with diameters of 5--10\,nm. 
    (\textbf{Mesoscopic}) Aggregates of clustered MNPs with effective diameters of 100--2000\,nm, treated as the primary species in the continuum model. 
    (\textbf{Continuum}) Homogenized slurry composed of multiple aggregates dispersed in the suspending fluid.
    }
    \label{fig:MNP_hierarchy}
\end{figure}

In equilibrium, MNPs assemble into mesoscale aggregates containing roughly 5\,\% solid phase ($\Phi_{MNP}\approx 0.05$), with the remaining volume occupied by entrained fluid~\cite{tesanovic_magnetic_2025}. This hierarchical organization—spanning single nanoparticles, aggregates, and ultimately the homogenized slurry—is depicted in Fig.~\ref{fig:MNP_hierarchy}. For modeling purposes, we treat these aggregates as the primary species, neglecting intra-aggregate restructuring and higher-order clustering. This simplification avoids additional length-scale coupling while retaining the essential features of magnetic and hydrodynamic interactions. The effective density of the aggregates can thus be expressed as
\begin{equation}
    \label{rho_agg}
    \rho_{a}(\Phi_{MNP}) = \rho_{MNP}\Phi_{MNP} + \rho_f(1-\Phi_{MNP}),
\end{equation}
where $\rho_{MNP}$ and $\rho_f$ denote the densities of the solid MNP and suspending fluid, respectively. Moreover as fluid is entrained in these aggregates, they will not produce the same magnetic susceptibility. This effective susceptibility for these aggregates can be approximated by clusters being fully connected, this microscopic perspective lends itself to various effective medium theories such as the Wiener's upper bound for anisotropic composites~\cite{wiener_abhandlungen_1912,torquato_random_2002}. From our proposed perspective for the MNPs to be part of an aggregate, they must be connected; therefore, the Wiener's upper bound is justified and thus utilized: 
    \begin{equation}
        \label{chi_agg}
        \chi_{a} = \Phi_{MNP}\chi_{MNP}+(1-\Phi_{MNP})\chi_f
    \end{equation}
Here, $\chi_{MNP}$ is the magnetic susceptibility of the MNP and $\chi_{f}$ is the magnetic susceptibility of the entrained fluid. Magnetic nanoparticles exhibit field-dependent magnetization ${\bf M}$ that saturates at high fields and no longer scales linearly with the applied magnetizing field, ${\bf H}=-\nabla\psi$~\cite{tesanovic_magnetic_2025}. Furlani and Ng~\cite{furlani_analytical_2006} captured this transition analytically by balancing the particle’s internal demagnetizing field with the applied field in a weakly susceptible fluid ($|\chi_f|\!\ll\!1$), yielding an effective susceptibility that crosses over from constant to $M_{sa}/|\nabla\psi|$. This saturation law is incorporated here to prevent unphysical force growth near strongly magnetized collectors.
\begin{equation}
    \label{demag}
    \chi_{a}^{sat}(\nabla\psi) = 
    \begin{cases} 
      \frac{3(\chi_a-\chi_f)}{(\chi_a-\chi_f)+3}, & |\nabla\psi| < \frac{(\chi_a-\chi_f)+3}{3\chi_a}M_{sa} \\
      \frac{M_{sa}}{|\nabla\psi|}, &  |\nabla\psi|\geq  \frac{(\chi_a-\chi_f)+3}{3\chi_a}M_{sa}  
    \end{cases}
\end{equation}
Here, $|\nabla\psi|$ is the applied magnetic field intensity and M$_{sa}$ is the saturation magnetization of our aggregates. These aggregates can be viewed as mesoscopic species that in turn when suspended in a fluid forms a slurry. This slurry can be treated as a continuum, i.e. considered macroscopically, and when captured by a surface it can be considered as the ``cake" in industrial applications.  The presence of these magnetically susceptible aggregates in the slurry as well as the suspending non-entrained fluid will perturb the magnetic field in a complex manner given their differences in magnetic susceptibility. Once again, we can leverage an effective medium theory like Hashin \& Shtrikman's lower bound for isotropic composites that views the aggregates as disconnected from each other by the suspending liquid~\cite{hashin_variational_1962,hashin_variational_1962-1,torquato_random_2002}. This perspective aligns with neglecting inter-aggregated agglomeration, hence we can write our continuum fluid's effective magnetic susceptibility as:
\begin{equation}
    \label{slurry_chi}
    \chi^{eff}_a (\tilde{c},\nabla\psi) = 3\left(\frac{3+\chi_{a}^{sat}}{3\tilde{c}+(3+\chi_a^{sat})(1-\tilde{c})}-1\right)
\end{equation}
Here, $\tilde{c} = v_a c$ where c is the number density of the aggregates and $v_a = (4\pi/3)r_a^3$ is the volume per aggregate. 
The maximum concentration of aggregates is taken as $\tilde{c}_{\max} = v_a c\, \Phi_{\max}$, with $\Phi_{\max} = 0.63$ corresponding to the random close packing of spheres. This value can be refined for systems with measured packing fractions or differing aggregate morphologies.

\subsubsection{Magnetochemical thermodynamics}
We propose the following free-energy functional for an inhomogeneous magnetic suspension:
\begin{equation}
    \label{GEnergy} 
    G = \int_V  \,d\textbf{r}\left\{g (\tilde{c},\nabla\psi)-\frac{1}{2}\mu_0 \nabla\psi\cdot\left(1+\chi_a^{eff}(\tilde{c},\nabla\psi)\right)\nabla\psi\right\}
\end{equation}
The first term g($\tilde{c},\nabla\psi$) is the enthalpy density, which depends on the concentration of magnetic aggregates and the \textbf{H}-field. This formulation allows the framework to be augmented for phase-field models. These effects will be neglected in the current model, i.e. g($\tilde{c},\nabla\psi$) $\approx$ g($\tilde{c}$), but their implications are discussed later. The second term subtracts the self-energy of the magnetic field using the approximation for the magnetic susceptibility introduced above. Although we do not consider chemically phase-separating mixtures, this term promotes magnetically induced pattern formation, as the free energy is lowered by co-locating regions of higher susceptibility, i.e. higher MNP concentration, in regions of larger magnetic fields. This phenomenon is analogous to electric-field induced pattern formation in dielectric materials with concentration-dependent permittivity~\cite{fraggedakis_dielectric_2020}. 

In order to predict how the slurry is modulated by the magnetic field, one needs a magnetostatic field equation modified by the concentration-dependent susceptibility. We derive a generalized, thermodynamically consistent Maxwell equation for the magnetic flux density, $\bf B$, by taking the variational derivative of the free energy functional and setting it equal to zero $\delta G/\delta \psi = 0$:
\begin{equation}
    \label{MLE}
\nabla\cdot{\bf B} =  -\mu_0\nabla\cdot\left(\left(1+\chi_a^{eff}(\tilde{c},\nabla\psi)\right)\nabla\psi+\frac{(\nabla\psi)^2}{2}\frac{\partial\chi_a^{eff}}{\partial\nabla\psi}\right) = 0
\end{equation}
Physically, this equation can be understood as prohibiting magnetic monopoles in the effective medium, where the ${\bf B}({\bf H})$ relation is defined variationally from the inhomogeneous free energy functional and includes a gradient correction as a result of saturation of the magnetic susceptibility. The generalized Maxwell equation, Eq.~\eqref{MLE}, for the magnetic scalar potential $\psi$ in an inhomogeneous medium with a nonlinear susceptibility is analogous to the generalized Poisson equation for the electrostatic potential in an inhomogeneous medium with a nonlinear polarizable dielectric. The underlying physics in these cases have been previously captured by free-energy functionals~\cite{fraggedakis_dielectric_2020,bazant_double_2011,garcia_thermodynamically_2004}.

In addition to Eq.~\eqref{MLE}, one needs the boundary conditions. Here, the boundary conditions reduce to the classical interfacial conditions of no magnetic flux and constant magnetic potential: (1) $\hat{\textbf{n}}\mathbf{\cdot}\textbf{B}_1 = \hat{\textbf{n}}\mathbf{\cdot}\textbf{B}_2$ and (2) $\psi_1=\psi_2$. Here $\hat{\textbf{n}}$ is the normal vector to the surface, $\textbf{B}_1 = -\mu_0\left(\left(1+\chi_a^{eff}(\tilde{c},\nabla\psi)\right)\nabla\psi+\frac{(\nabla\psi)^2}{2}\frac{\partial\chi_a^{eff}}{\partial\nabla\psi}\right)$ is our slurry's magnetic field at the interface, $\textbf{B}_2$ is the classical magnetic field at the interface in the wire or other solid magnetic surfaces, and $\psi_i$ is the magnetic scalar potential at the interface in region i. 

Additionally, from the free energy functional, one can derive the magnetochemical potential, $\mu = \delta G/\delta c$:
\begin{equation}
    \label{gen_mu}
    \mu = \frac{\delta G}{\delta c}= k_B T \ln\left(\tilde{c}\right)+\mu^{ex}(\tilde{c})-\frac{\mu_0v_a}{2}\left(\nabla\psi\cdot\nabla\psi\right)\frac{\partial \chi_a^{eff} (\tilde{c},\nabla\psi)}{\partial \tilde{c}}
\end{equation}

\noindent where $\mu^{ex}$ is the excess chemical potential which accounts for non-idealities, $k_B$ is the Boltzmann constant, and T is the temperature. Here we implement the Carnahan-Starling formulation for the excess chemical potential, which accounts for the steric effects of random sphere packing~\cite{carnahan_equation_1969,bazant_towards_2009}:
\begin{equation}
    \label{CS_mu}
    \frac{\mu^{ex}}{k_BT} = \frac{\delta g^{CS}}{\delta c} = \frac{\tilde{c}\left(8-9\tilde{c}+3\tilde{c}^2\right)}{\left(1-\tilde{c}\right)^3}
\end{equation}

\noindent In full for this work, the magnetochemical potential can be written as:
\begin{equation}
    \label{ChemPot}
    \mu = \frac{\delta G}{\delta c}= k_B T \ln\left(\tilde{c}\right)+k_B T \frac{\tilde{c}\left(8-9\tilde{c}+3\tilde{c}^2\right)}{\left(1-\tilde{c}\right)^3} -\frac{\mu_0v_a}{2}\left(\nabla\psi\cdot\nabla\psi\right)\frac{\partial \chi_a^{eff} (\tilde{c},\nabla\psi)}{\partial \tilde{c}}
\end{equation}

\noindent The magnetochemical potential is critical for understanding how the individual aggregates accumulate as well as the thermodynamic pressure they induce in the hydrodynamics. The thermodynamic force density (\textbf{f}), can be obtained by calculating the thermodynamic forces on an individual particle ($-\nabla\mu$) and multiplying it by the number density of the particles ($c$). This approach comes from the Gibbs-Duhem relationship and has been discussed in previous works~\cite{groot_non-equilibrium_1962,batchelor_brownian_1976,bazant_towards_2009}:
\begin{equation}
    \label{GDR}
    -\mathbf{f} = c\nabla\mu = \nabla P_0 + \nabla P_m
\end{equation}

\noindent Here, $\nabla P_0$ represents the osmotic pressure that balances the concentration gradient and $\nabla P_m$ represents the magnetic pressure that balances the forces due to the variations in the magnetic susceptibility due to the concentration gradient: 
\begin{equation}
    \label{os_p}
    \nabla P_0 = k_BT \left(1+\tilde{c}^2\frac{8-2\tilde{c}}{(1-\tilde{c})^4}\right)\nabla c
\end{equation}
\begin{equation}
    \label{mag_p}
    \nabla P_m = -\frac{\mu_0c}{2}\nabla\left((\nabla\psi)^2\frac{\partial \chi_a^{eff}}{\partial c}\right)
\end{equation}

\subsection{Mass Transport}
As we want to view the fluid as a homogenized slurry, we need two conservation of mass equations, one for the fluid and one for the magnetic nanoparticles:
\begin{equation}
    \label{com}
    \frac{\partial\rho_{eff}}{\partial t} + \nabla\cdot\left(\rho_{eff}\textbf{v}\right) = 0
\end{equation}

\noindent with \textbf{v} being the fluid velocity. From the proposed modeling framework, the slurry can be viewed as a homogeneous suspension, allowing the effective density to be directly calculated:
\begin{equation}
    \label{rho_eff}
    \rho_{eff}(\tilde{c}) = \rho_{a}\tilde{c} + \rho_f(1-\tilde{c})
\end{equation}

\noindent Beyond the conservation of mass for the slurry, the aggregates must uphold species conservation:
\begin{equation}
    \label{cos}
    \frac{\partial c}{\partial t} + \nabla\cdot\left(c\textbf{v}_a-\frac{Dc}{k_BT}\nabla\mu\right) = 0
\end{equation}

\noindent Here D is the diffusivity of the aggregates and \textbf{v}$_a$ is the velocity of the aggregates. This perspective is employed as our slurry contains magnetic particles that exist as mesoscopic aggregates and not as its own fluid phase. Therefore, we need a force balance on the individual aggregates:
\begin{equation}
    \label{FB}
    \frac{3}{2}\rho_ac\left(\frac{\partial \textbf{v}_a}{\partial t} + \textbf{v}_a\cdot\nabla\textbf{v}_a\right) = \rho_ac\frac{\textbf{v}-\textbf{v}_a}{\tau_a}f(Re_a)+c\frac{\mu_0}{2}(\chi_a-\chi_f)\nabla\left(\nabla\psi\right)^2
\end{equation}

\noindent where $\tau_a=2\rho_a r_a^2/(9\eta_f)$ is the aggregate's response time, f(Re$_a$) captures the non-idealities in the drag coefficient; however as in all the cases in this work Re$_a = 2\rho_ar_a\textbf{v}_0/\eta_f < 1$, one can consider f(Re$_a$)$\approx$ 1~\cite{maxey_equation_1983,bagchi_effect_2003,kasbaoui_clustering_2019}. For simplicity we will neglect the inertial terms for the magnetic nanoparticles, which reduces the equation down to:
    \begin{equation}
        \label{FB_NI}
        \textbf{v}_a = \textbf{v} + \frac{r_a^2\mu_0}{9\eta_f}\left(\chi_a-\chi_f\right)\nabla\left(\nabla\psi\right)^2
    \end{equation}

\noindent This means our conservation of species equation can be written as:
    \begin{equation}
        \label{cos_red}
        \frac{\partial c}{\partial t} + \nabla\cdot\left(c\textbf{v} + c\frac{r_a^2\mu_0}{9\eta_f}\left(\chi_a-\chi_f\right)\nabla\left(\nabla\psi\right)^2-c\frac{D}{k_BT}\nabla\mu\right) = 0
    \end{equation}

\subsection{Hydrodynamics}
Lastly, we need to apply the conservation of momentum to our slurry. For simplicity, we will assert the slurry to be a Newtonian fluid without bulk viscosity due to isotropic expansion. This approach results in a modified Navier-Stokes momentum equation with osmotic and magnetic pressure:
    \begin{equation}
        \label{MNS-Eqn}
         \rho_{eff}\left(\frac{\partial\textbf{v}}{\partial t} + \textbf{v} \cdot \nabla \textbf{v}\right) = -\nabla P + \nabla \cdot \left(\eta_{eff}\left[\left\{\nabla\textbf{v}+\left(\nabla\textbf{v}\right)^t\right\}-\frac{2}{3}(\nabla\cdot\textbf{v})\mathbf{I}\right]\right)-\nabla P_0 -\nabla P_m
    \end{equation}

\noindent Here, we have neglected gravitational effects as prior models have done \cite{chen_dynamic_2012}. Additionally as we have a suspension of aggregates, the viscosity will be a function of the concentration:
    \begin{equation}
    \label{eq:eta}
        \eta_{eff}(\tilde{c}) = \eta_f\left(1-\tilde{c}/\tilde{c}_{max}\right)^{-2}
    \end{equation}

This formulation follows prior theories for the viscosity of colloidal suspensions~\cite{brady_rheological_1993,mewis_colloidal_2013}. Future refinements to the proposed model could account for the non-Newtonian nature of the deviatoric stress or potential complexity from the conservation of angular momentum in super-paramagnetic suspensions.

\section{Results and Discussion}

To explore the predictive capabilities of the continuum framework developed herein, we simulate the accumulation of superparamagnetic nanoparticles near magnetized cylindrical wires under varying magnetic field strengths, flow velocities, and matrix orientations. For all simulations, we assert $\chi_f = 0$, consistent with the weak magnetic susceptibility of water and many biological buffers, thereby simplifying the effective medium theory without loss of generality. All physical and geometric parameters used in the simulations are summarized in the Appendix \ref{sec:appendix} (Table~\ref{tab:parameters}). 

The resulting concentration fields $\tilde{c}(\mathbf{x}, t)$, shown in Figures~\ref{fig:result_steady} and \ref{fig:result_temp}, reveal both steady-state and time-dependent morphologies that emerge from the nonlinear coupling between magnetophoretic transport, hydrodynamic advection, and field-mediated phase interactions. These patterns reflect the fundamental physics encoded in the free energy formulation: particle accumulation perturbs the local field, thereby altering the driving forces for further deposition, and resulting in an inherently self-regulating process. We analyze the captured particle mass as a function of the Mason number ($\mathrm{Mn}$), which compactly expresses the balance between magnetic and viscous forces in HGMS. As shown in Figure~\ref{fig:Mason}, this reveals three distinct regimes of capture.

While the current model captures bulk transport and saturation effects with high fidelity, it does not resolve fine-scale morphological instabilities such as dendrite formation. Such structures have been observed in Euler–Lagrange simulations of magnetophoretic capture~\cite{chen_dynamic_2012} and may be interpreted as interfacial instabilities driven by spatial gradients in effective susceptibility, analogous to dielectric-induced phase separation in electrochemical systems~\cite{bazant_theory_2013, fraggedakis_dielectric_2020, gao_interplay_2021}. Extensions of the present framework to include higher-order interfacial terms, such as gradient energy or curvature-driven fluxes, would enable the emergence of such features within a thermodynamically consistent phase-field model.

\subsection{Data Representation and Scaling}

All concentration fields are reported as normalized aggregate volume fractions, defined by $(\tilde{c} - \tilde{c}_0)/\tilde{c}_{\mathrm{max}}$, where $\tilde{c}_0$ denotes the initial uniform concentration and $\tilde{c}_{\mathrm{max}}$ corresponds to the random close-packing limit of aggregates, taken as $\Phi_{\mathrm{max}} = 0.63$.

To facilitate comparison across disparate operating conditions—particularly variations in magnetic field strength $B_{\text{0}}$ and flow velocity $v_{\text{0}}$—all colormaps are scaled relative to their local maxima. For each ($B_{\text{0}}$,$v_{\text{0}}$) combination, the color axis is capped at 10\% of the peak concentration obtained in that simulation, and this limit is applied identically across both matrix orientations (parallel and orthogonal). This consistent scaling prevents visual saturation in low-capture regimes while enhancing contrast in dilute regions, thereby exposing subtle spatial features otherwise obscured by absolute concentration differences. The resulting representations preserve the qualitative morphology of accumulation patterns while maintaining quantitative interpretability.

\subsection{Steady-State Accumulation Patterns}

\begin{figure}[h!]
    \centering
    \includegraphics[width=0.75\linewidth]{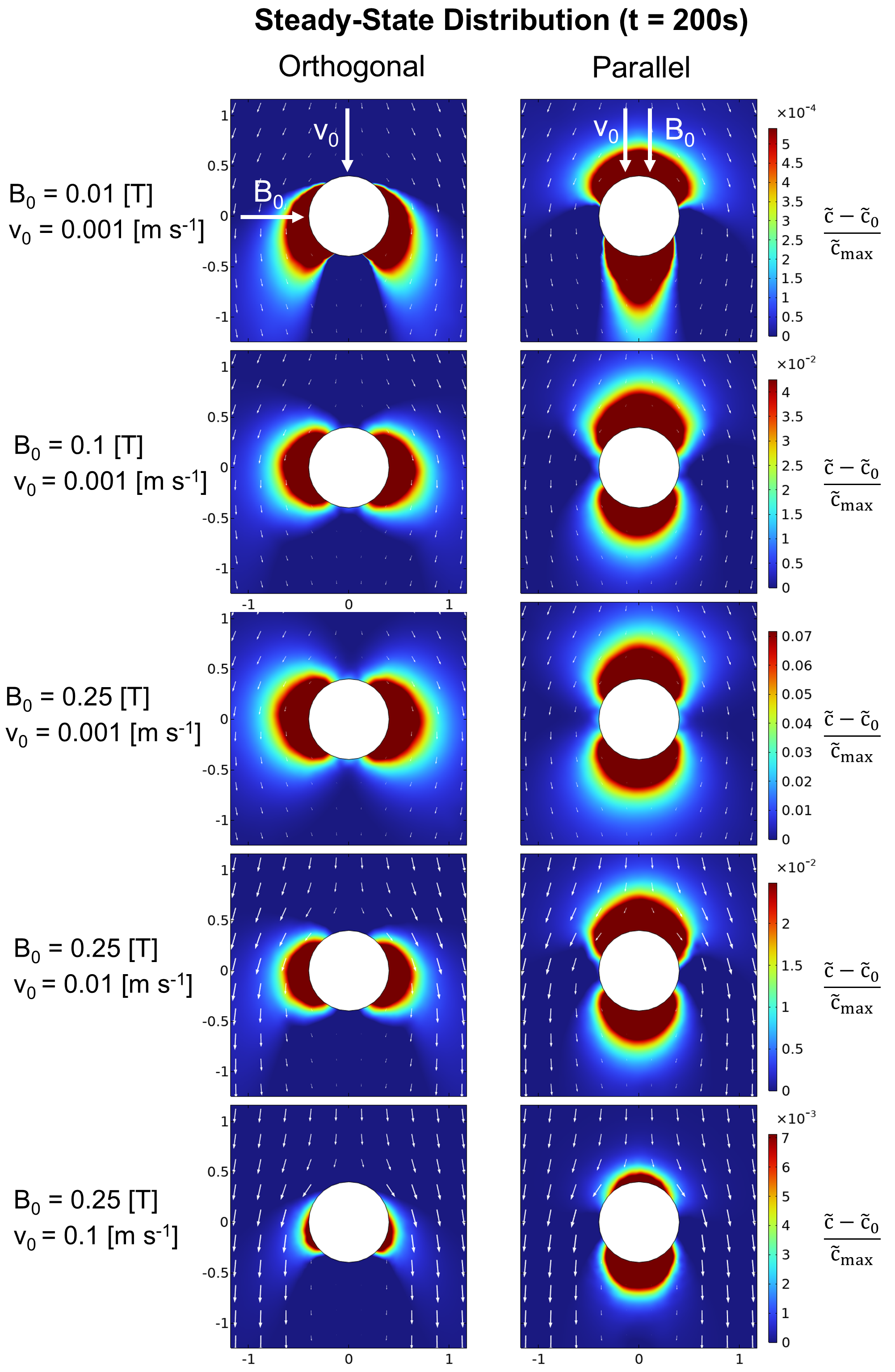}
    \caption{Steady-state magnetic particle concentration fields at $t = 200$\,s for orthogonal (left) and parallel (right) wire configurations, shown across increasing magnetic field strengths [$B_{\text{0}} = 0.01$, 0.1, 0.25\,T] and flow velocities [$v_{\text{0}} = 0.001$, 0.01, 0.1\,m\,s$^{-1}$]. Colormaps show normalized concentration $(\tilde{c} - \tilde{c}_0)/\tilde{c}_{\mathrm{max}}$, scaled to 10\% of the peak value for each condition to enhance contrast. Arrows indicate the imposed flow direction $\mathbf{v}$. Spatial axes are in [mm].}
    \label{fig:result_steady}
\end{figure}

Figure~\ref{fig:result_steady} illustrates the normalized steady-state concentration fields $\tilde{c}(\mathbf{x})$ after $t = 200$\,s of accumulation under varying magnetic field strengths $B_0$, inlet flow velocities $v_{\text{0}}$, and matrix orientations. The results highlight the interplay between magnetophoretic drift and hydrodynamic advection in shaping the spatial organization of magnetic aggregates.

\newpage

At low field strength ($B_{\text{0}} = 0.01$\,T) and low flow velocity ($v_{\text{0}} = 0.001$\,m\,s$^{-1}$), particle accumulation remains weak, with local concentrations rising by less than 1\% of the random close-packing limit. The magnetic body force is too small to overcome hydrodynamic drag; and, most particles remain entrained in the bulk flow. Nevertheless, the resulting concentration fields display nontrivial structure driven by the interplay of magnetophoretic drift and convective transport. In the orthogonal configuration, lateral magnetic deflection generates crescent-shaped plumes on the upstream face, which are advected downward into asymmetric wing-like tails. In the parallel case, particles are focused axially into a broad, flattened accumulation at the front stagnation zone, followed by a narrow, shielded wake extending downstream. These dilute but organized morphologies reflect the directional coupling of field and flow and are consistent with accumulation patterns observed in similar HGMS experiments~\cite{hu_dynamic_2022}.

As the magnetic field strength increases to $B_{\text{0}} = 0.1$ and $0.25$\,T at fixed inlet velocity ($v_{\text{0}} = 0.001$\,m\,s$^{-1}$), pronounced accumulation zones emerge adjacent to the wire, driven by intensified magnetic gradients. Local concentrations rise to approximately 42\% and 71\% of the random close-packing limit, respectively, signaling the onset of near-jamming conditions. In the orthogonal configuration, particles are laterally deflected into a crescent-shaped region upstream of the wire, elongated axially by the flow and slightly truncated along the flanks where shear dominates. This asymmetry reflects the transverse orientation of field and flow, which promotes lateral focusing but limits surface retention under drag. In contrast, the parallel configuration produces axially symmetric deposits at the front and rear stagnation zones, where magnetophoretic drift is either aligned with or opposed by the flow, enabling stable accumulation at both poles of the collector.

At fixed magnetic field strength ($B_{\text{0}} = 0.25$\,T), increasing the flow velocity from $v_{\text{0}} = 0.001$ to 0.01 and 0.1\,m\,s$^{-1}$ reduces peak accumulation from 71\% to 24\% and 7\% of the random close-packing limit, respectively. The capture zone is compressed into a narrow boundary layer adjacent to the wire, reflecting the inverse scaling of residence time and capture thickness with velocity. As advection outpaces magnetophoretic drift, fewer particles are redirected toward the wire before exiting the domain.

Overall, the simulations reveal that capture morphology emerges from a force balance between magnetophoretic drift, and hydrodynamic drag, modulated by concentration-dependent field shielding. The relative orientation of $\mathbf{B}$ and $\mathbf{v}$ dictates whether these forces act cooperatively, producing axially symmetric deposits (parallel case), or competitively, generating crescent-shaped plumes prone to lateral stripping (orthogonal case). As concentrations approach the random close-packing limit ($\Phi \approx 0.63$), steric stresses and permeability loss compress the accumulation zone into thin boundary layers, further amplifying drag–field competition. These morphologies closely resemble experimental observations and prior predictive models of wire-based HGMS~\cite{hu_dynamic_2022, franzreb_magnettechnologie_2003, chen_dynamic_2012, choomphon-anomakhun_simulation_2017}. 

\subsubsection*{Dimensionless Phase Diagram for Field-Induced Particle Capture}

Total capture was quantified as a steady excess \emph{inventory} per unit depth:  
\[
M_{\mathrm{cap}} = N^{\mathrm{on}}(t_f) - N^{\mathrm{off}}(t_f),
\]
where 
\[
N(t) = \int_{\Omega_f} c(\mathbf{x},t)\,dA
\]
is the integrated particle concentration in the flow cross–section at time \(t\) (detailed in Section~\ref{sec:appendix}).  
This definition measures the net inward mass accumulated in the presence of the magnetic field relative to a reference run without the field.

To compare capture efficiency across different field strengths \(B_{0}\) and flow velocities \(v_{0}\), we organize the data using the Mason number,
\[
\mathrm{Mn} \;\sim\; \frac{\eta_f\,v_{0}}{\mu_{0}\,\chi\,B_{0}^{2}\,r_w},
\]
which compares viscous drag to magnetophoretic driving and delineates the crossover from drift–dominated to advection–dominated transport.  
Here, \(r_w\) denotes the collector (wire) radius, and Mn is defined using far–field parameters, omitting \(\mathcal{O}(1)\) geometric prefactors for canonical scaling.  
Using \(r_w\) as the length scale for field variation,
\[
\big|\nabla\!\left(|\mathbf{B}|^{2}\right)\big| \;\sim\; \frac{B_{0}^{2}}{r_w},
\quad \text{where} \quad |\mathbf{B}|^{2} \equiv \mathbf{B} \!\cdot\! \mathbf{B}.
\]

\begin{figure}[h]
    \centering
    \includegraphics[width=1\linewidth]{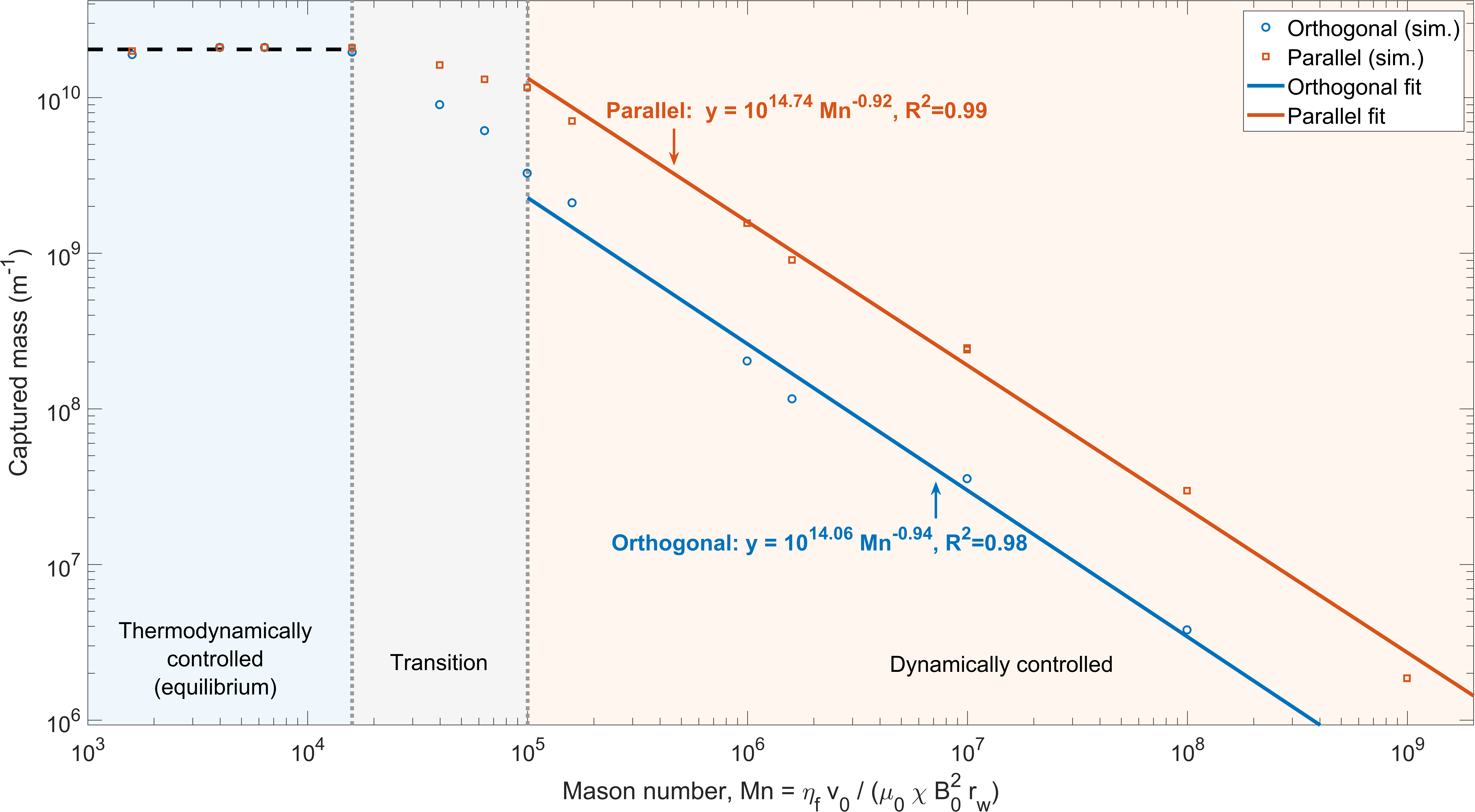}
    \caption{
    Dimensionless phase diagram for field–induced particle capture: Captured mass versus Mason number for orthogonal (blue) and parallel (orange) field–flow configurations. 
    The Mason number is defined with far–field parameters as
    \(\mathrm{Mn}=\eta_f v_{0}/(\mu_{0}\chi B_{0}^{2} r_w)\),
    where \(r_w\) is the wire (collector) radius and \(\chi\equiv\chi_a\).
    Three regimes are visible: thermodynamically controlled (\(\mathrm{Mn}\ll10^{4}\)), transition (\(\mathrm{Mn}\sim10^{4}\)), and dynamically controlled (\(\mathrm{Mn}\gg10^{5}\)).
    Solid lines are power–law fits; the fitted exponents are
    \(-0.94\) (orthogonal) and \(-0.92\) (parallel) with \(R^{2}>0.98\).
    Points at very low velocity (\(v_{0}=10^{-4}\,\text{m\,s}^{-1}\)) approach a saturation plateau.
    The y–axis reports the cumulative inward particle mass per unit depth.
    }
    \label{fig:Mason}
\end{figure}

Figure~\ref{fig:Mason} presents this data in the form of a \emph{dimensionless phase diagram} for field-induced particle capture, with \(M_{\mathrm{cap}}\) plotted versus \(\mathrm{Mn}\) for both parallel and orthogonal field–flow configurations. Across six orders of magnitude in \(\mathrm{Mn}\) obtained by systematically varying \(B_0\) and \(v_0\), three distinct regimes emerge. At low Mason numbers, \(\mathrm{Mn} \ll 10^{4}\), the captured mass is independent of both \(\mathrm{Mn}\) and field orientation. This thermodynamically controlled limit arises when magnetophoretic drift is strong enough to bring nearly all accessible particles to the collector surface within their residence time; therefore, hydrodynamic shear plays no role and capture is instead limited by near–surface crowding and concentration–dependent magnetic shielding (Eq.~\ref{MLE}). Once the local concentration approaches the jamming threshold, further reductions in \(v_{0}\) or increases in \(B_{0}\) produce no measurable gain.

As \(\mathrm{Mn}\) increases toward \(\sim 10^{4}\), a transition regime appears in which hydrodynamic effects begin to influence capture and field orientation becomes relevant. In the parallel configuration, converging streamlines at the upstream stagnation zone enhance residence time and reduce tangential loss; whereas in the orthogonal configuration, transverse drift delivers particles into high–shear flanks, promoting detachment and streamline bypass in the boundary layer. This onset of orientation dependence marks the breakdown of purely thermodynamic control.

For \(\mathrm{Mn} \gg 10^{5}\), the system enters a dynamically controlled regime governed by the competition between advection and magnetophoretic drift. Here, the captured mass decreases as \(M_{\mathrm{cap}} \propto \mathrm{Mn}^{-0.92}\) for the parallel configuration and \(M_{\mathrm{cap}} \propto \mathrm{Mn}^{-0.94}\) for the orthogonal configuration, with \(R^{2} > 0.98\). The nearly parallel scaling lines indicate similar underlying transport mechanisms, while the constant vertical offset between them reflects persistent differences in near–surface flow topology. Here, the partial drift–flow alignment in the parallel case sustains higher capture rates; where as, the transverse drift-flow alignment in the orthogonal case feeds particles into regions of strong shear lowering net accumulation. Framing the results in terms of this dimensionless phase diagram reveals clear physical boundaries between thermodynamically controlled, transitional, and dynamically controlled regimes. This finding enables \textit{in silico} optimization and guides rational HGMS design.

\subsection{Temporal Build-Up, Propagation, and Field Shielding}

\begin{figure}[h!]
    \centering
    \includegraphics[width=0.6\linewidth]{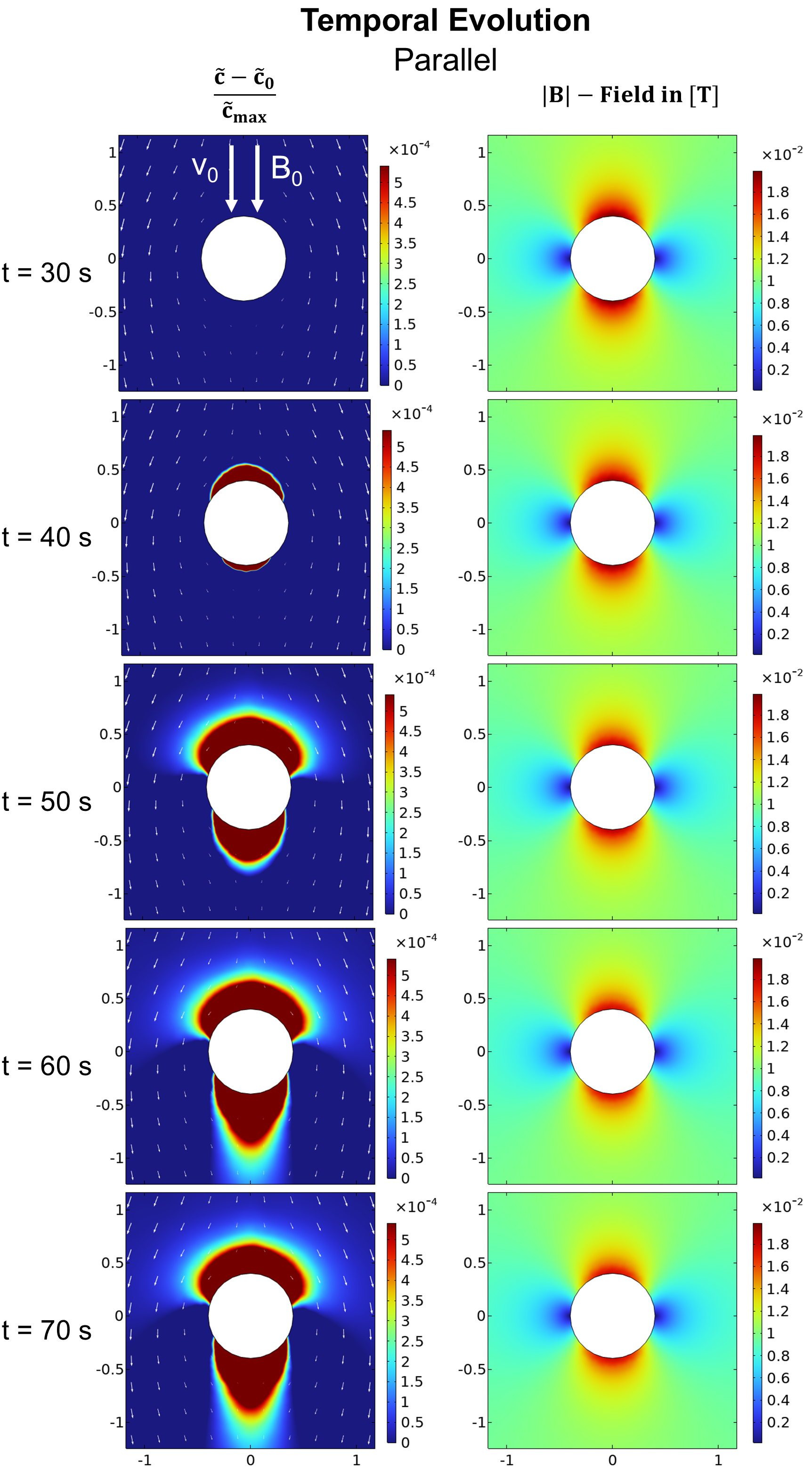}
    \caption{Temporal evolution of magnetic particle accumulation (left) and magnetic field magnitude (right) in the parallel configuration at $B_{\text{0}} = 0.01$\,T and $v_{\text{0}} = 0.001$\,m\,s$^{-1}$. Concentrations are normalized by $c_{\mathrm{max}}$, the random close-packing limit. Magnetic field intensity [T] shows a $\sim$5\% decrease from $t = 30$\,s to $t = 70$\,s, with the high-gradient zone retreating toward the wire surface (within $\sim$0.1\,mm).}
    \label{fig:result_temp}
\end{figure}

Figure~\ref{fig:result_temp} shows the coupled evolution of particle concentration (left panels) and magnetic field magnitude (right panels) in the parallel configuration at $B_{\text{0}} = 0.01$\,T and $v_{\text{0}} = 0.001$\,m\,s$^{-1}$. At $t = 30$\,s, before significant deposition, the concentration field remains uniform while the magnetic field exhibits the undisturbed dipolar pattern of the bare wire. By $t = 40$–50\,s, a localized accumulation forms at the upstream stagnation point and elongates downstream into a narrow finger aligned with the flow. The deposit thickens over time approaching a quasi-steady morphology by $t = 60$–70\,s with sharp concentration gradients confined to a thin boundary layer.

Simultaneously, the magnetic field distribution exhibits measurable feedback: comparison of early and late frames reveals a $\sim$5\% decrease in peak field intensity at the wire surface and a contraction of the high-gradient zone to within $\sim$0.1\,mm ($\sim$13\% of the wire diameter). This shielding arises from the concentration-dependent susceptibility of the captured particles, which partially screens the applied field and limits further accumulation. Such self-limiting behavior—emerging from the coupling between transport and field distortion—is a defining feature of nonlinear magnetic capture dynamics.

\newpage

\section{Conclusion and Outlook}

We introduced a thermodynamically consistent continuum framework for magnetic‐particle transport and capture in high–gradient fields that couples (i) a modified magnetostatic equation with concentration–dependent susceptibility, (ii) advection–diffusion with magnetochemical driving forces, and (iii) a momentum balance closed by osmotic and magnetic body stresses. Derived from a free–energy functional, the model predicts directional accumulation, shear–induced detachment, boundary–layer confinement, and partial field shielding without empirical shutoff rules.

Numerical simulations reproduce typical HGMS morphologies across field strengths, flow rates, and matrix orientations allowing for rationalization of performance trends through force–balance scalings. 
When recast into a dimensionless phase diagram based on the Mason number, the captured mass organizes into three regimes: a thermodynamically controlled limit at low $\mathrm{Mn}$ where capture saturates and is insensitive to flow–field orientation; a transition regime where hydrodynamic shear begins to influence accumulation; and a dynamically controlled regime at high $\mathrm{Mn}$ where advection competes with magnetophoretic drift and orientation strongly affects capture. 
Departures from scaling at the lowest velocities confirm saturation from near–surface crowding and magnetic shielding. 
These regime boundaries provide mechanistic guidance for tuning field strength, flow rate, and collector orientation to balance capture efficiency and throughput.

Beyond steady state, the free–energy formulation can be extended naturally to predict transient breakthrough curves, track the evolution of pressure drop, and simulate particle capture in three–dimensional matrix designs. The same formulation can be extended to resolve interfacial instabilities (e.g. dendritic growth) by adding gradient–energy terms. Additional refinements could incorporate non-Newtonian rheology. Furthermore, this formulation could enable inverse design and optimal control by differentiating through the governing equations. More broadly, the framework is applicable to processes that steer or concentrate magnetic particulates—from bioseparations to drug delivery—and provides a rigorous foundation for data-assimilated digital twins that link microscopic physics to process-scale design and operation.

\section*{Acknowledgments}
We are grateful to Leonard Maier, and J. Pedro de Souza for the helpful discussions and feedback. M.T. \& S.B. acknowledge support from Technical University of Munich's Global Incentive Fund. D.M.M. \& M.Z.B. acknowledge support from the Center for Enhanced Nanofluidic Transport 2 (CENT$^2$), an Energy Frontier Research Center funded by the U.S. Department of Energy (DOE), Office of Science, Basic Energy Sciences (BES), under award \# DE-SC0019112. D.M.M. also acknowledges support from the National Science Foundation Graduate Research Fellowship under Grant No. 2141064. Any opinions, findings, and conclusions or recommendations expressed in this material are those of the author(s) and do not necessarily reflect the views of the National Science Foundation.

\newpage

\renewcommand{\figurename}{Figure}
\renewcommand{\theequation}{S\arabic{equation}} 
\renewcommand{\thefigure}{S\arabic{figure}} 
\renewcommand{\thetable}{S\arabic{table}}  
\renewcommand{\labelenumii}{\arabic{enumi}.\arabic{enumii}}
\renewcommand{\labelenumiii}{\arabic{enumi}.\arabic{enumii}.\arabic{enumiii}}
\renewcommand{\labelenumiv}{\arabic{enumi}.\arabic{enumii}.\arabic{enumiii}.\arabic{enumiv}}
\setcounter{figure}{0}

\section{Appendix}\label{sec:appendix}
\subsection{Methods}

\subsubsection{Numerical simulation}
Numerical simulations were carried out in COMSOL Multiphysics\textsuperscript{\textregistered}~6.2 using built‑in physics interfaces to solve the coupled magnetic, flow, and transport equations derived in the theory section. Three modules were combined: (i) Magnetic Fields, No Currents for magnetostatics, (ii) Laminar Flow for incompressible Stokes flow, and (iii) Stabilized Convection–Diffusion for advective–diffusive transport with streamline stabilization to handle high Péclet numbers.

\begin{figure}[h]
    \centering
    \includegraphics[width=0.55\linewidth]{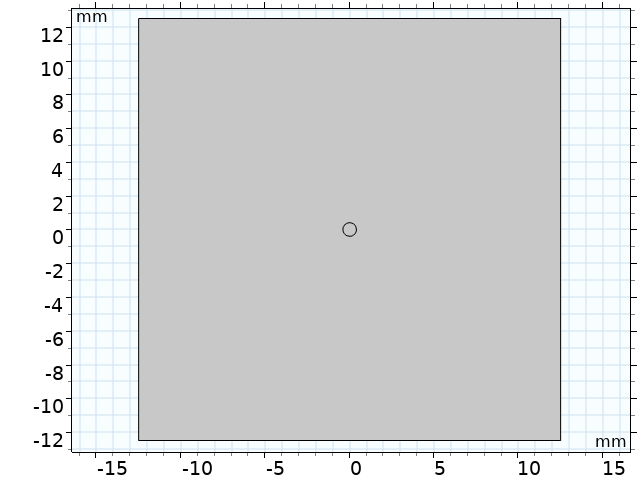}
    \caption{Computational domain: magnetizable cylindrical collector (center) in a nonmagnetic fluid. Outer boundaries set the magnetic scalar potential; internal boundaries enforce no‑flux and no‑slip conditions.}
    \label{fig:domain}
\end{figure}

The computational domain shown in Figure~\ref{fig:domain} was a two‑dimensional cross‑section containing a single magnetizable cylinder embedded in a larger square fluid region, chosen to minimize boundary effects. External magnetic fields were imposed via scalar potential boundaries, while transport used no‑flux conditions at solid surfaces. Flow boundaries employed no‑slip walls and an open outlet (zero gauge pressure); inflow velocity was ramped smoothly to avoid startup transients.

Triangular finite elements with local refinement around the wire resolved steep field and concentration gradients; mesh independence was verified by comparing capture flux across refinements. Time integration used an implicit BDF scheme with adaptive stepping and strict tolerances. Nonlinear solves employed a segregated strategy, iterating magnetic, flow, and transport fields until convergence. All simulations were two‑dimensional, and results are reported per unit depth. Post‑processing and normalization were performed in COMSOL.

For clarity, the fixed material properties and geometric dimensions defining the model domain are listed in Table~\ref{tab:parameters}; only the applied field strength and flow velocity were varied in the simulations.

\subsubsection{Captured MNP Mass}
In 2D we compute capture as an \emph{excess inventory per unit depth}.  
Let \(\Omega_f\) be the fluid domain and \(c(\mathbf x,t)\) the volumetric particle concentration
\([\mathrm{m}^{-3}]\). The instantaneous inventory is
\[
N(t)\;=\;\int_{\Omega_f} c(\mathbf x,t)\,\mathrm dA \qquad [\mathrm{m}^{-1}] .
\]
Two otherwise identical simulations are run (same mesh, \(v_0\), \(B_0\), numerics): a \emph{field-on}
case \(N^{\mathrm{on}}(t)\) and a \emph{field-off} baseline \(N^{\mathrm{off}}(t)\).
The captured mass per unit depth reported in Fig.~\ref{fig:Mason} is the steady excess inventory
\[
M_{\mathrm{cap}} \;=\; N^{\mathrm{on}}(t_f) \;-\; N^{\mathrm{off}}(t_f),
\]
with \(t_f\) chosen after \(N^{\mathrm{on}}(t)\) plateaus
(Implemented in COMSOL via \emph{Surface Integration} of \(c\) over \(\Omega_f\), i.e. a domain integral in 2D.).
Note that for the orthogonal orientation at $v_0=0.1~\mathrm{m\,s^{-1}}$ with $B_0=0.01$ and $0.1~\mathrm{T}$, the computed excess mass was slightly negative; these outliers were omitted from Fig.~\ref{fig:Mason}, as they most likely reflect numerical limitations of the finite $25~\mathrm{mm}$ domain at high flow rates (short residence time) rather than a physical effect.

\begin{table}[h]
\centering
\caption{Fixed physical and geometric parameters used in the continuum HGMS model. Notation follows the theoretical framework in Section~\ref{sec:theory}.}
\begin{tabular}{llll}
\hline
\textbf{Parameter} & \textbf{Unit} & \textbf{Description} & \textbf{Value} \\
\hline
$r_a$ & nm & Aggregate radius (representative particle size) & $1000$ \\
$\chi_{\mathrm{MNP}}$ & -- & Magnetic susceptibility of MNP material & $6$ \\
$T$ & K & Temperature (thermal energy scale) & $300$ \\
$\rho_{\mathrm{MNP}}$ & kg\,m$^{-3}$ & Density of magnetic nanoparticles & $5170$ \\
$\rho_f$ & kg\,m$^{-3}$ & Density of suspending fluid (water) & $1000$ \\
$\Phi_{\max}$ & -- & Maximum packing fraction (random close packing) & $0.63$ \\
$\Phi_{\mathrm{MNP}}$ & -- & Solid fraction of MNP within aggregates & $0.05$ \\
$\mu_0$ & H\,m$^{-1}$ & Magnetic permeability of free space & $1.257\times10^{-6}$ \\
$\mu_w$ & H\,m$^{-1}$ & Magnetic permeability of water & $1.257\times10^{-6}$ \\
$\mu_{\text{iron}}$ & H\,m$^{-1}$ & Magnetic permeability of collector (iron) & $50\,\mu_0$ \\
$M_{sa}$ & A\,m$^{-1}$ & Saturation magnetization of aggregates & $3.6\times10^{5}$ \\
$c_0$ & -- & Initial particle concentration at inlet (normalized) & $0.01\,c_{\max}$ \\
$r_w$ & mm & Wire (collector) radius & $0.8$ \\
Domain size & mm & Outer boundary dimensions of 2D simulation & $15 \times 15$ \\
\hline
\end{tabular}
\label{tab:parameters}
\end{table}

\newpage

\bibliography{library}

\end{document}